\documentclass[a4paper,fleqn,usenatbib]{mnras}
\usepackage[T1]{fontenc}
\usepackage{ae,aecompl}
\usepackage{amsmath,graphicx,color,ulem} 
\usepackage{amsfonts,amssymb}

\usepackage[british]{babel}

\hypersetup{draft}

\newcommand{\turb}{_{\rm t}}  
\newcommand{\hot}{_{\rm hot}}     
\newcommand{\vect}[1]{\boldsymbol{#1}} 
\newcommand{\bvec}[1]{\boldsymbol{#1}}

\newcommand{\xyavg}[1]{\left<#1\right>_{xy}}

\newcommand{\rsub}[2]{#1_{\mathrm{#2}}}

\newcommand\deriv[2]{\frac{\partial #1}{\partial #2}}

\definecolor{cceblue}{RGB}{0,0,128}

\definecolor{fagreen}{rgb}{0,0.5,0}

\definecolor{purple}{rgb}{0.5,0,.5}

  \newcommand{\cm}{\,{\rm cm}}

  \newcommand{\erg}{\,{\rm erg}}
  \newcommand{\g}{\,{\rm g}}

  \newcommand{\kms}{\,{\rm km\,s^{-1}}}

    \newcommand{\mkG}{\,\umu{\rm G}}
  \newcommand{\K}{\,{\rm K}}
  \newcommand{\kpc}{\,{\rm kpc}}
  \newcommand{\pc}{\,{\rm pc}}

  \newcommand{\Gyr}{\,{\rm Gyr}}
  
\defcitealias{GSFSM13}{Paper~I}
\defcitealias{GSSFM13}{Paper~II}

\title[Magnetic effects]{The supernova-regulated ISM -- VI. 
Magnetic effects}
\author[C.~C.~Evirgen, F.~A.~Gent, A.~Shukurov, A.~Fletcher and P.~J.~Bushby]{
C.~C.~Evirgen,$^{1}$\thanks{
E-mail: c.c.evirgen@newcastle.ac.uk
} 
F.~A.~Gent,$^{2}$ A.~Shukurov,$^{1}$ A.~Fletcher$^{1}$ and P.~J.~Bushby$^{1}$\\
$^{1}$School of Mathematics, Statistics and Physics, Newcastle University, 
Newcastle upon Tyne, UK, NE1 7RU\\
$^{2}$ReSoLVE Centre of Excellence, Department of Computer Science, Aalto 
University, PO Box 15400, FI-00076 Aalto, Finland
}
\date{Accepted --. Received --; in original form --}
\pubyear{2018}

\volume{{\rm in press}}
\begin{document}
\pagerange{\pageref{firstpage}--\pageref{lastpage}}
\maketitle
\label{firstpage}

\begin{abstract}
We explore the effect of magnetic fields on the vertical distribution and 
multiphase structure of the supernova-driven  interstellar medium (ISM) in simulations that admit 
dynamo action.
As the magnetic field is amplified to become dynamically significant,
gas becomes cooler and its distribution in the disc becomes more homogeneous.
We attribute this to magnetic quenching of vertical
velocity, which leads to a decrease in the cooling length of hot gas.
A non-monotonic vertical distribution of the large-scale magnetic field strength,
with the maximum at $|z|\approx 300$~pc causes a downward
pressure gradient below the maximum which acts against outflow driven by SN explosions,
while it provides pressure support above the maximum.
\end{abstract}

\begin{keywords}
galaxies: magnetic fields, galaxies: kinematics and dynamics, ISM: evolution, 
MHD, turbulence
\end{keywords}

\section{Introduction}
In the dynamics of the interstellar medium 
(ISM), 
the role of magnetic fields is most often discussed in the contexts of the 
pressure support of the galactic gas layer \citep[][and references 
therein]{B87,BC90,FS01}, galactic winds, especially with cosmic rays 
\citep{BMKV91,BMKV93,EZBMCRG08}, and star formation \citep{PBKML11,C12}. 
There is no clear consensus about the dynamical significance of 
magnetic fields in the ISM as a whole.
For example, \citet{Avillez05,HJMBM12,SILCC1} suggest a modest 
or negligible magnetic field contribution,
while \citet[][and references therein]{F01,Cox05}
 argue for a significant dynamical role.
The exploration of interstellar magnetic fields and their effects upon the 
structure and dynamics of the multi-phase ISM is complicated by 
difficulties in observing them, for example, by the absence of 
observational estimates for magnetic field strength in the hot gas.

One way to clarify the picture is to employ increasingly powerful and
realistic numerical simulations. Numerical simulations have
explored many aspects of the ISM, covering a variety of physical 
effects on a broad range of scales from sub-parsec to kiloparsec.
Each numerical model must exclude some physical processes at 
relevant length and time scales, but each helps to 
clarify the significance of particular physical effects. Relevant numerical models 
include those of \citet{Tomisaka98,Korpi99a,WN99,Avillez00, WN01, WMN02, 
	AB04, Balsara04, Slyz05, Avillez05, MacLow05, Joung06, WN07, AB07, 
	Gressel08, Joung09, Piontek09, Wood10, HJMBM12, GSFSM13, BGE15, LFSR15, 
	SILCC1, Pardi17,Yadav17} and \citet{CGO18}.

\citet{Evirgen16} discussed the effects of the multi-phase ISM  
structure on the mean and random galactic
magnetic fields in a numerical
simulation of the supernova-driven multi-phase ISM. Here, we explore 
the effects of the magnetic field on the ISM including its multi-phase 
structure, gas outflow and the force balance. We identify several effects 
which are rather unexpected and yet, with hindsight, physically compelling.

The structure of the paper is as follows. In Section~\ref{sec:Model}, we 
describe the numerical model. In Section~\ref{sec:vert_str}, we discuss
the vertical structure of the simulated ISM, focusing on the
changes in the vertical velocity and thermodynamic structure due to magnetic fields.
The dynamical significance of the magnetic field is the subject of Section~\ref{sec:gas_pressure}.
This is followed by discussion of the role of magnetic fields in the vertical force balance,
with reference to disc structure and vertical velocity in Section~\ref{sec:force}.
Section~\ref{sec:bz} focuses on the vertical profile of the magnetic field,
and its relation to the other constituents of the ISM.
The conclusions are summarised in Section~\ref{sec:Conclusions}.
\section{Simulations of the SN-driven ISM}\label{sec:Model}
A local Cartesian box, $1\times1\kpc^2$ in size horizontally and extending to 
$1\kpc$ on each side of the galactic mid-plane, is placed at a galactocentric 
radius of $8\kpc$ \citep[][hereafter, \citetalias{GSFSM13}]{GSFSM13}. The local 
Cartesian coordinates 
$(x,y,z)$ correspond, respectively, to the cylindrical polar coordinates 
$(r,\phi\,z)$ with the $z$-axis aligned with the galactic angular velocity. 
Parameters are representative of the Solar neighbourhood, but with rotation 
double the rate in the Milky Way to accelerate magnetic field amplification by 
the dynamo, as discussed in 
\citet[][hereafter, \citetalias{GSSFM13}]{GSSFM13} and \citet{Evirgen16}; the 
numerical model used here is denoted B2$\Omega$ in \citet{Gent12}. Supernova 
sites, where thermal and 
kinetic energies are injected into the gas, are distributed randomly in time 
and space at the occurrence frequency of the Solar neighbourhood. The numerical 
resolution is 4\,pc in each direction; with such a grid spacing, it is possible to reproduce the known 
expansion laws of supernova (SN) remnants from the Sedov--Taylor to the late 
snowplough phases \citepalias{GSFSM13,GMKSH18}.   
Differential rotation is implemented using the shearing periodic boundary
conditions in the radial ($x$) direction. The system of non-ideal, fully compressible and nonlinear magnetohydrodynamic 
(MHD) equations is solved assuming the equation of state of an ideal monatomic 
gas. The simulations use the ISM module of the Pencil 
Code\footnote{https://github.com/pencil-code}. The momentum equation includes 
velocity shear due to galactic differential rotation, the 
Coriolis force, viscous stress, kinetic energy injection by SNe, and the 
Lorentz force. A fixed gravity field is due to the stellar mass and the 
dark matter following \cite{Kuijken89}. The energy equation includes viscous and 
Ohmic heating and thermal energy injected by the SNe. Radiative cooling is 
parametrised using the cooling functions of \citet{Sarazin87} and 
\citet{Wolfire95}, and photoelectric heating follows \citet{Wolfire95}.
The cooling rate is truncated at $T=100\K$ to avoid numerically intractable
gas densities; the heat diffusion is enhanced to ensure that gas clouds produced
by thermal instability are fully resolved at the working numerical resolution
\citepalias[{further details can be found in} Appendix~B of][]{GSFSM13}. 
Self-gravity is neglected given the relatively low gas densities in the 
simulated ISM, $n\lesssim10^2\cm^{-3}$ in terms of the number density. The 
induction equation is solved in terms of the vector potential to ensure the 
solenoidality of the simulated magnetic field. The detailed form of the
equations can be found in \citetalias{GSFSM13}.
\begin{figure}
	\centering
	\includegraphics[width=0.99\linewidth]{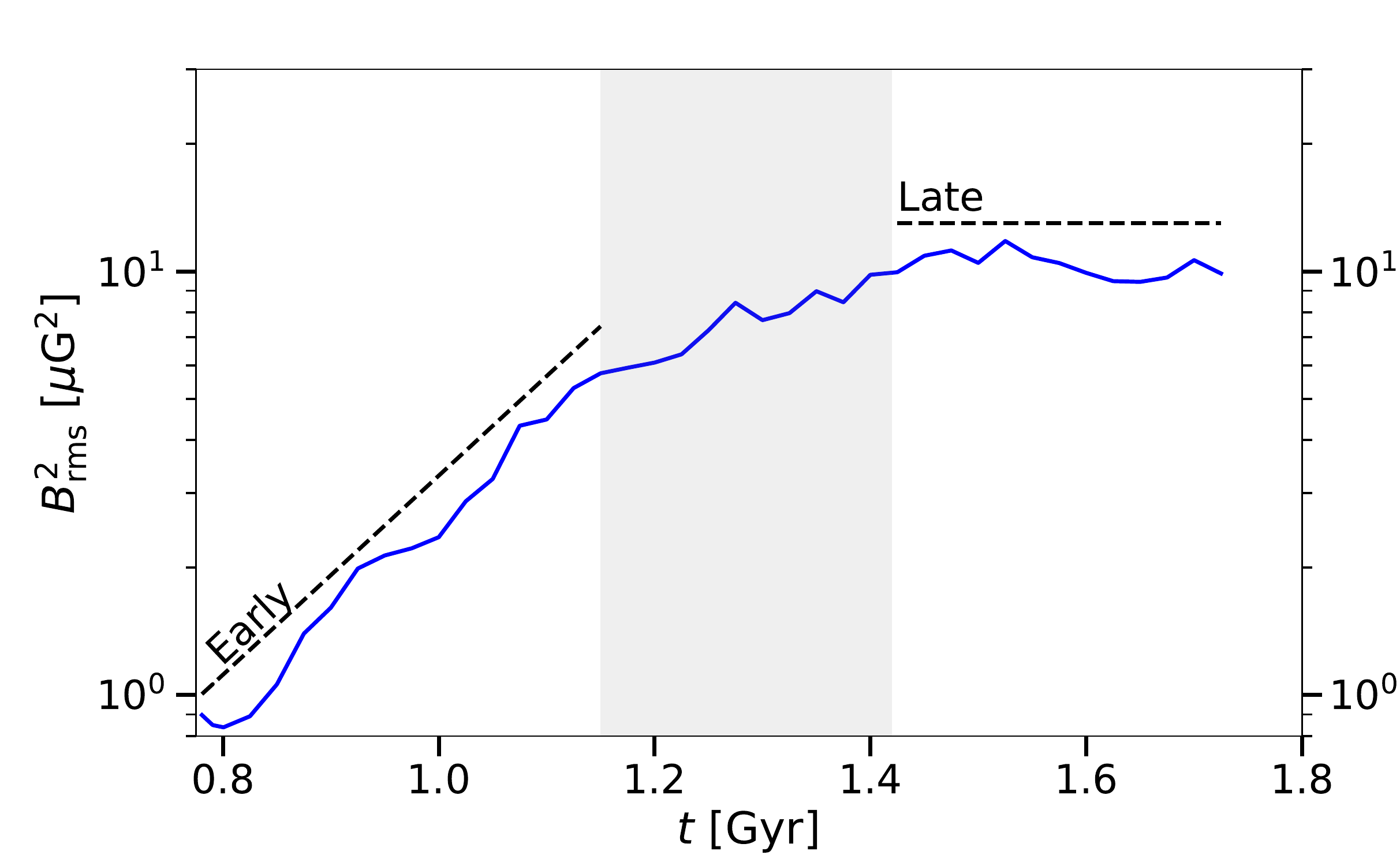}
\caption{Evolution of the mean-squared magnetic field strength. During 
the Early (kinematic) stage of the dynamo, the initially weak field grows 
exponentially, with a mean slope indicated by the dashed line. The dynamo is in 
a statistically steady state during the Late stage. The shaded time interval 
represents the transition between these phases.
}
	\label{fig:B2O_bsq}
\end{figure}

Figure~\ref{fig:B2O_bsq} shows the time evolution of the volume-averaged 
magnetic energy density $B^2$ and introduces the Early and Late stages of the 
model evolution. During the Early stage, $0.78\lesssim t \lesssim1.15\Gyr$, the 
magnetic field is too weak to influence the flow or perturb 
the thermodynamic structure of the ISM; the dynamo is therefore kinematic, 
which explains the exponential growth in the magnetic field strength. Next 
follows a transitional stage ($1.15 \lesssim t \lesssim 1.42\Gyr$) when the 
growth of the field slows down as the Lorentz force gradually becomes 
strong enough to exert a dynamical influence upon the flow. The system 
settles to a statistically steady state in the Late stage, $t \gtrsim1.42\Gyr$, 
where the energy density of the magnetic field is comparable to 
that of the random motions and thermal energy. By comparing the system during 
the Early and Late stages (the former with a negligible magnetic field, the 
latter with a dynamically significant field) it is therefore possible to 
identify the effects of magnetic fields on the ISM.
We stress that the Early stage represents a hydrodynamical statistically
steady state of the system, whereas the Late stage is an MHD steady state. 
Wherever appropriate, we present results for the transitional stage despite its 
transient nature, since it may be observable in high-redshift galaxies and 
to illustrate the continuity of the adjustments between the Early and Late 
stages.
\begin{figure*}
	\includegraphics[width=0.99\linewidth]{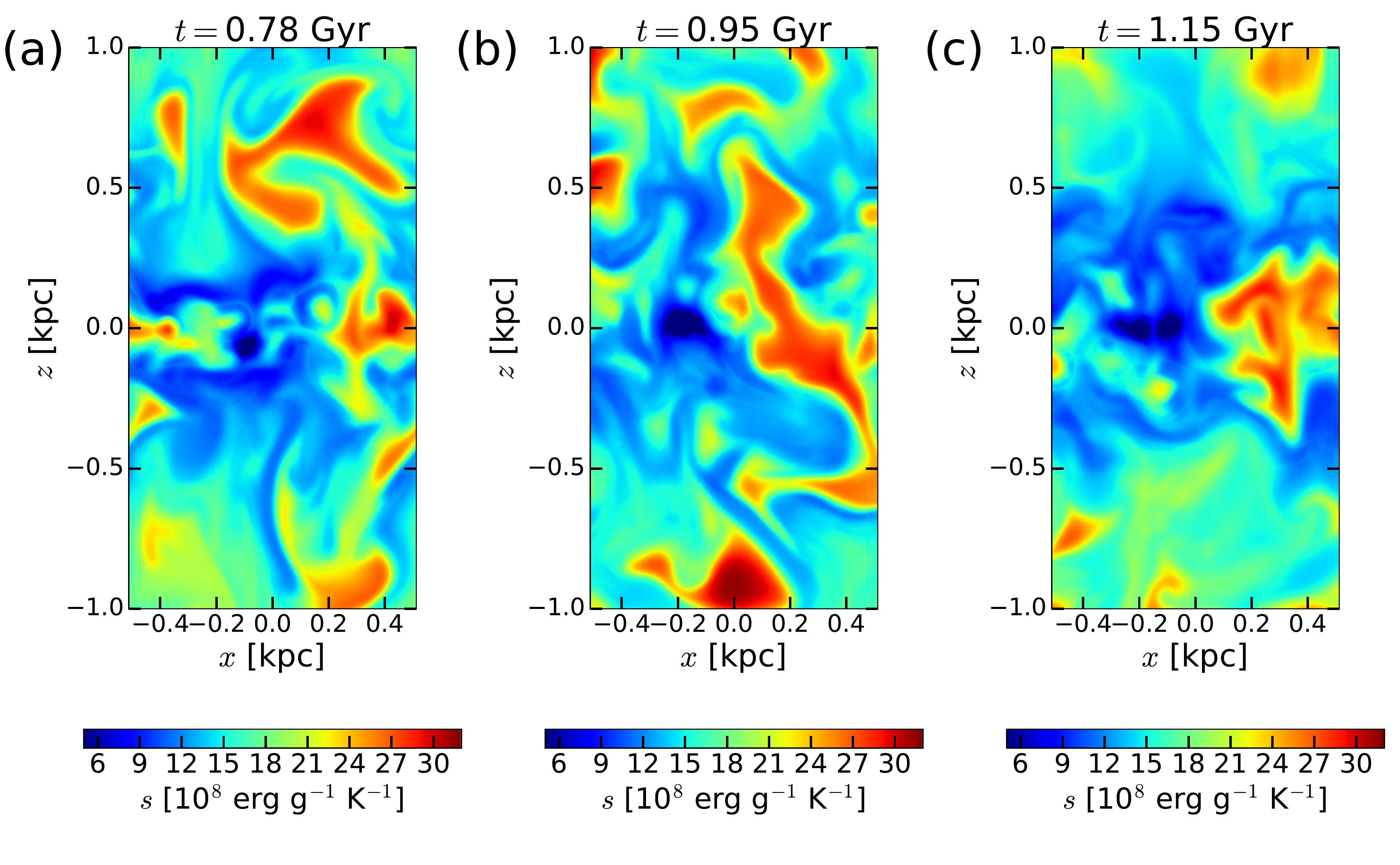}
	\includegraphics[width=0.99\linewidth]{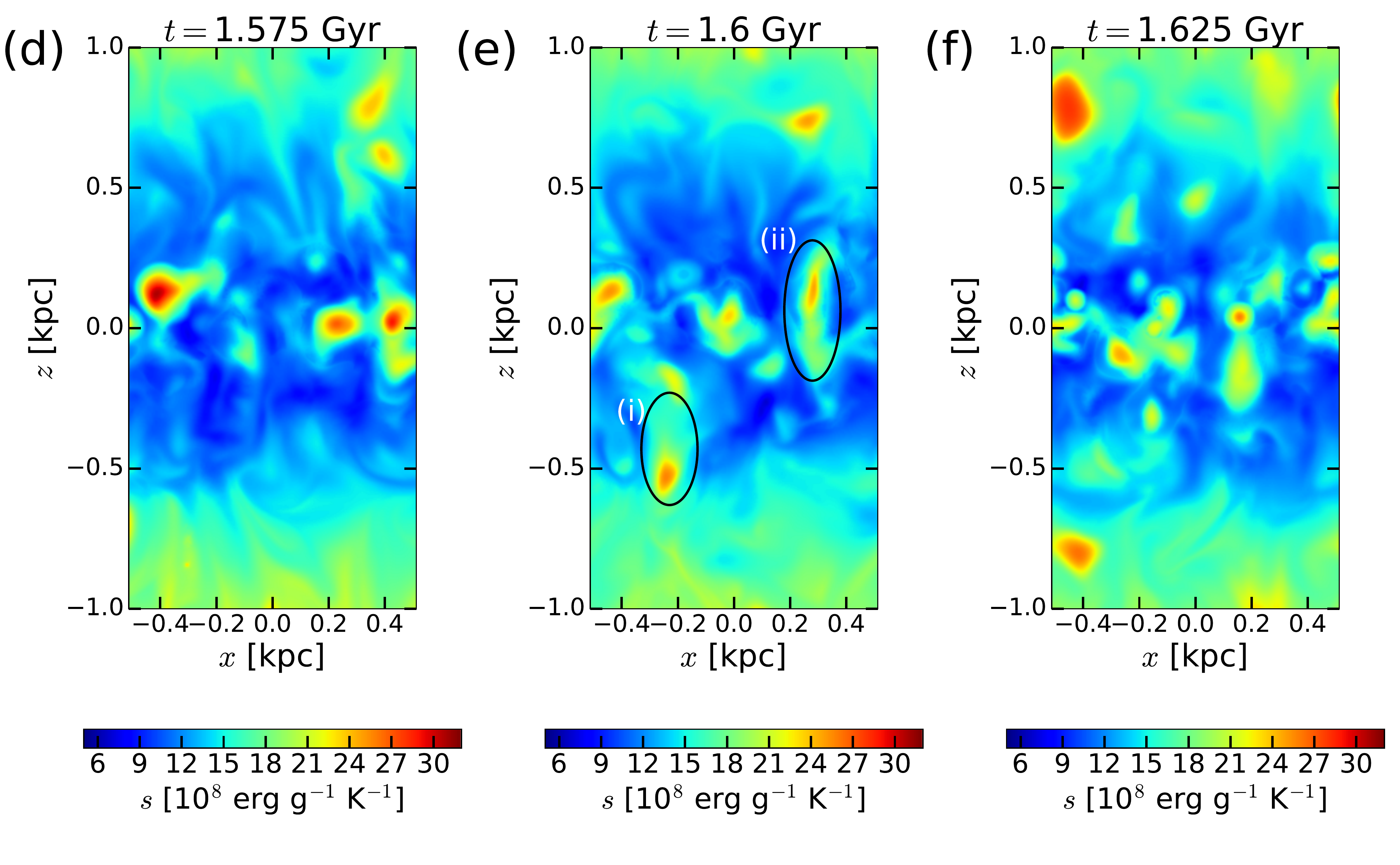}
	\caption{Specific entropy in the $(xz)$-plane, corresponding to 
	$(r,{z})$ of the cylindrical frame, with the large-scale velocity shear 
	in the $y$-direction: \textbf{(a)}, \textbf{(b)} and {\textbf{(c)}} are 
	from the Early stage whereas panels \textbf{(d)}, \textbf{(e)} and 
	\textbf{(f)} represent the Late stage. The time after the start of 
	the simulation is given at the top of each panel and entropy colour 
	bars are provided at the bottom. It is evident that the system is more 
	homogeneous in the Late stage, with a lower abundance of the hot gas. The 
	structures labelled (i) and (ii) in panel~(e) are discussed in the text.}
	\label{fig:sn_fx}
\end{figure*}
\section{Vertical structure of the ISM}\label{sec:vert_str}
It is generally accepted that magnetic fields can affect the structure and dynamics 
of the interstellar medium. However, the magnetic effects are still not fully understood, with 
a number of important questions still unresolved. In this section, we present detailed
comparison of the simulated ISM in its Early and Late stages to identify the 
ways in which magnetic fields affect this system. 
\subsection{{Changes to the vertical distribution of specific entropy and gas density}}
\label{sec:vert_str_ent}
Figure~\ref{fig:sn_fx} shows the specific entropy distribution for a few 
representative snapshots, at various stages of evolution of the simulated ISM.
The specific entropy {of the interstellar gas is defined by}
\begin{equation}\label{eq:ent}
s = c_V\left[\ln\frac{T}{T_0}-(\gamma-1)\ln\frac{\rho}{\rho_0}\right],
\end{equation}
where $c_V$ is the specific heat capacity at constant volume, $T$ 
and $\rho$ are the gas temperature and density, with the reference values 
$T_0=1\K$ and $\rho_0=1\g\cm^{-3}$, and $\gamma=5/3$ is the adiabatic index. 
Specific entropy is quoted in the text in units of 
$10^8\erg\g^{-1}\K^{-1}$. 

\noindent The top panels of Fig.~\ref{fig:sn_fx} are taken from the Early stage (in which
the magnetic field is dynamically negligible); the bottom panels represent the Late stage 
(when the magnetic field becomes dynamically important). 
Specific gas entropy $s$ is colour coded, with red 
corresponding to hot and dilute gas, and blue to cooler and denser gas. 
{Blue colours represent} the 
warm phase of the simulated ISM ($4.4<s<23.2$, 
$500<T<5\times 10^5\K$, $10^{-26}<\rho<10^{-24}\g\cm^{-3}$ for the specific 
entropy, temperature and density, respectively); the cold gas occupies a 
small fraction of the volume and is hardly visible in this representation. 

In the Early stage, large hot gas structures are widespread and many of them
span a large part of the domain. In the Late stage, the hot
structures are typically smaller and rounder. This suggests that the magnetic field
tends to drive the system towards a more homogeneous gas distribution, given that 
any qualitative changes arise from the evolution from a hydrodynamically
steady state to a magnetohydrodynamically steady state. Some
indications of this behaviour can already be seen in panel\,\textbf{(c)} of
Fig.~\ref{fig:sn_fx}, which corresponds to the end of Early stage (when local
magnetic fields can already be dynamically important); the regions of hot gas
already seem to be less extensive than those at earlier times.

Panel\,\textbf{(e)} of Fig.~\ref{fig:sn_fx} contains two features, labelled (i) and (ii), that demonstrate the 
reduced ability of hot gas to expand in the Late stage. They show hot structures
formed close to the mid-plane and rising to larger $|z|$ producing little
disturbance in the surrounding gas.
\begin{figure}
	\centering
	\includegraphics[width=0.95\linewidth]{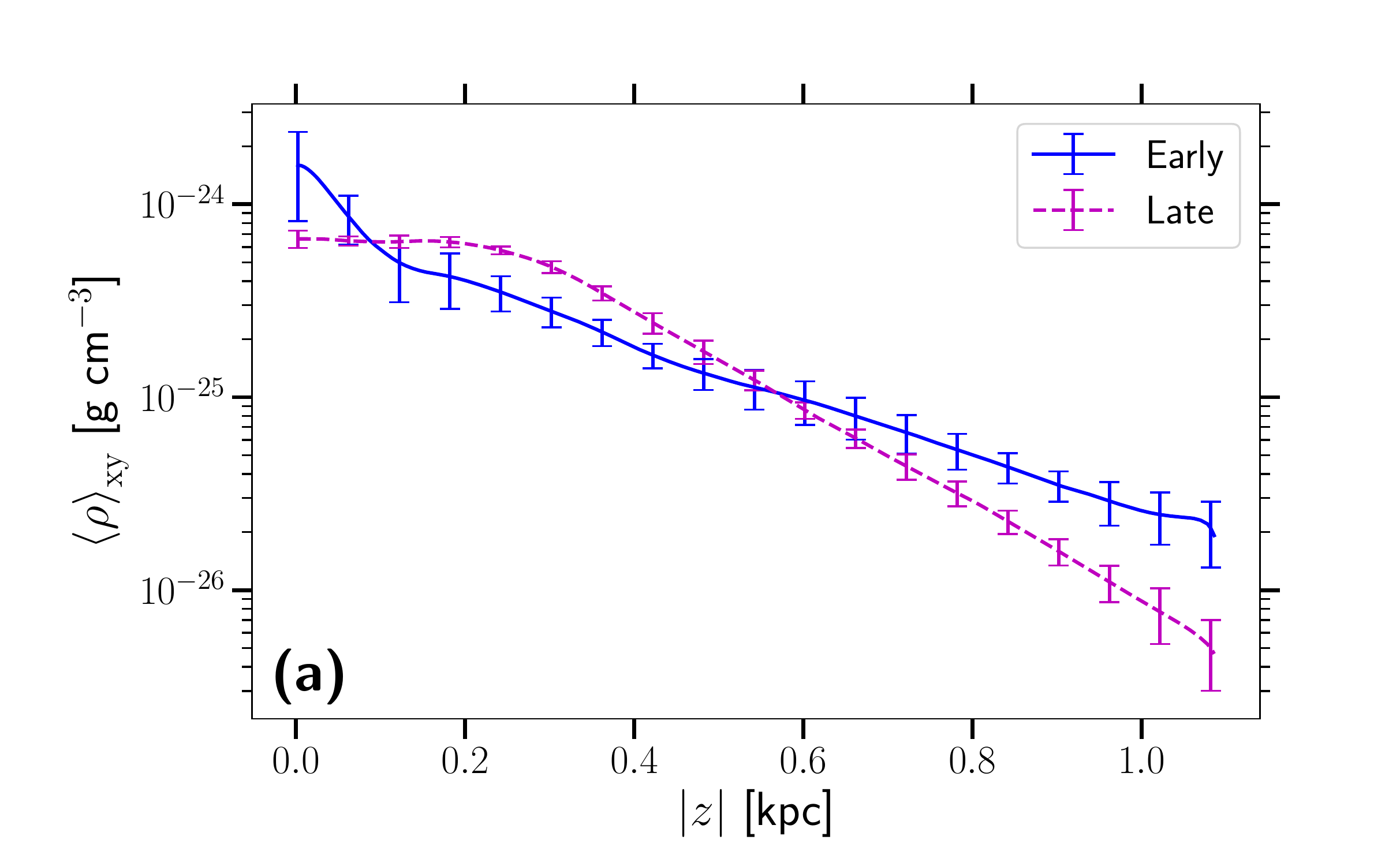}
	\includegraphics[width=0.95\linewidth]{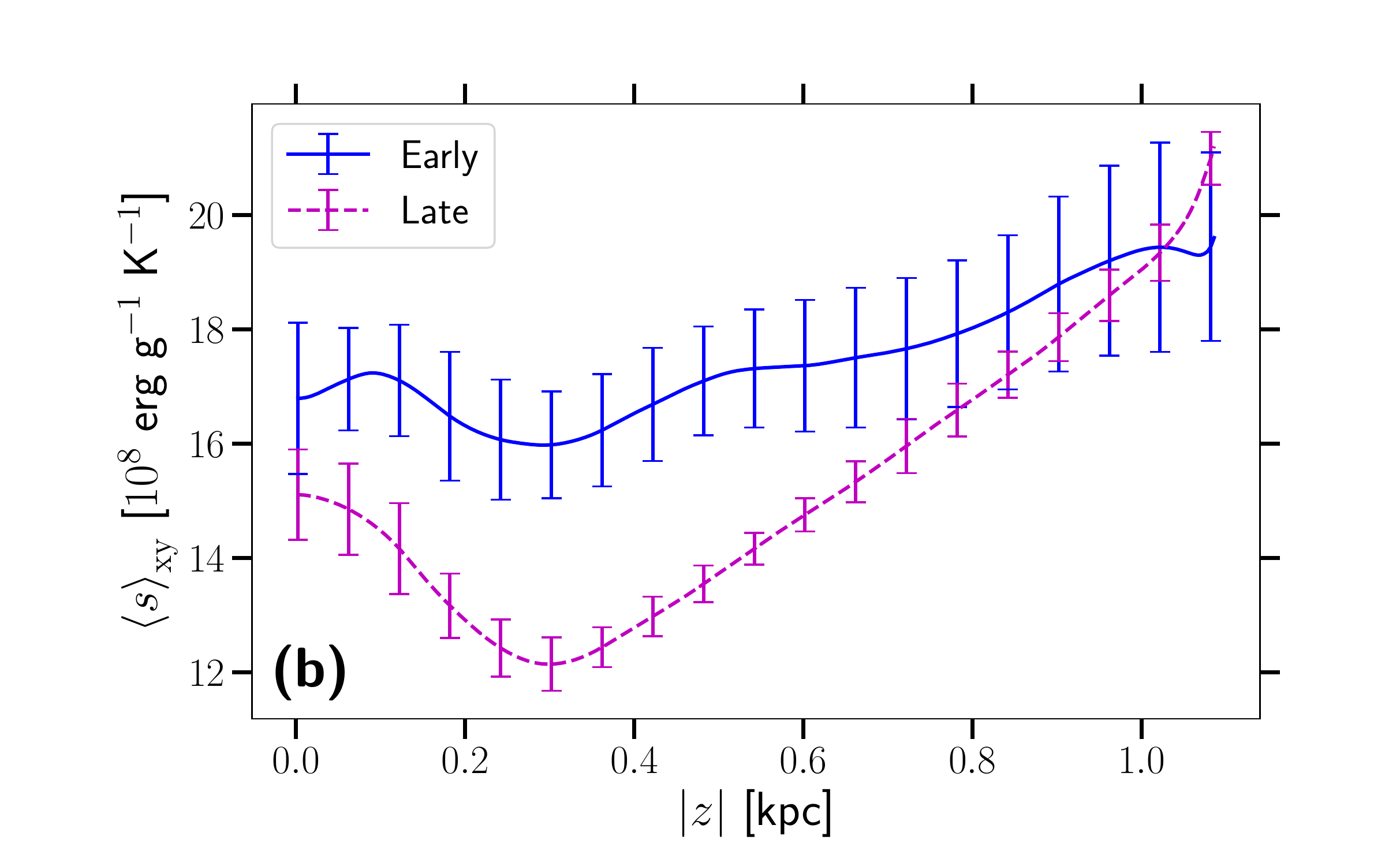}
\caption{The horizontally averaged \textbf{(a)}~gas density 	
and \textbf{(b)}~specific entropy versus distance to the mid-plane, $|z|$, further 
averaged over time for the Early (solid, blue) and Late (dashed, magenta) 
stages of the evolution.}
\label{fig:B2O_rho}
\end{figure}

Figure~\ref{fig:B2O_rho} 
provides an alternative view of the 
gas structure showing the horizontally averaged gas 
density and specific entropy, as functions of distance $z$ to the mid-plane, averaged over time for
the Early \textbf{(17 snapshots)} and Late \textbf{(13 snapshots)} stages separately. During the Early stage, the mean gas density 
is maximum at the mid-plane, decreasing rapidly with height within $100\pc$ of 
the midplane and then more gradually.
In the Late stage, there is a clear flattening of the density profile for 
$|z|\lesssim0.3\kpc$, but a steeper density gradient for 
$|z|\gtrsim0.3\kpc$. 
The specific entropy is everywhere lower 
in the Late stage than it is in the Early stage, which is consistent with the 
apparent reduction in the abundance of hot gas. 
A pronounced minimum in the specific entropy 
at around  $|z|\simeq0.3\kpc$ is notable -- this is the same height at 
which the mean density gradient changes.

\citet[][their Figure~9\textbf{a} and Table~4]{HJMBM12} present vertical profiles of gas 
density in MHD simulations of the SN-driven ISM where magnetic field is not 
generated self-consistently by the dynamo action, as in our model, but imposed to 
be initially independent of $x$ and $y$ and scale with the initial gas density as 
$n^{1/2}(z)$. This model shows that the gas distribution is 
rather insensitive to the strength of the imposed magnetic field, when it varies
between $0$ and $10~\mkG$ at the midplane.
The vertical gas density 
profile in the Early stage of our Model is similar to the density profiles of 
\citet{HJMBM12}. Thus, the ISM containing magnetic field produced by the system 
itself in a self-consistent manner via dynamo action is very different from that of an imposed field.

\subsection{Enhanced cooling of hot gas in a magnetised ISM}
\noindent The effects described above are noticeable over a scale of hundreds
of parsecs. However, both the smaller size of hot gas structures, seen in Fig.~\ref{fig:sn_fx},
and the decrease in entropy -- particularly close to $|z|=300$~pc, seen in
Fig.~\ref{fig:B2O_rho}\,\textbf{b}, suggest that hot gas is affected by the magnetic
field at smaller scales as well.
\begin{figure}
  \includegraphics[width=0.9\linewidth]{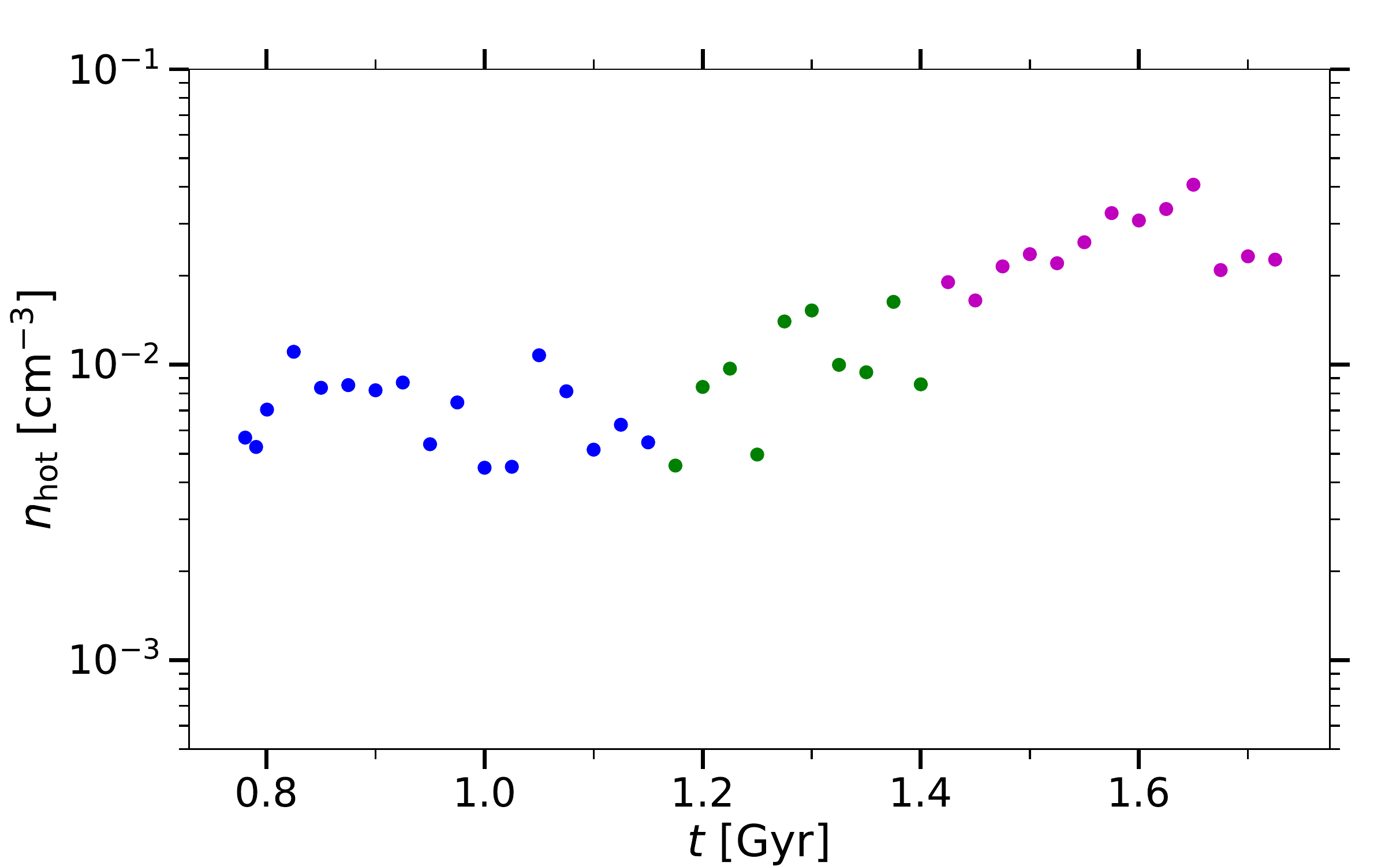}
  \caption{The average number density of the hot gas as the Model evolves.
	Blue, green and magenta circles refer to values from the Early,
	Transitional and Late stages, respectively.}
  \label{fig:nhot}
\end{figure}
We find that fractional volume of the hot gas decreases from 20--25\% in the Early stage
to 1--5\% in the Late stage. \citet{MSHRBF18} also find differences in the topology of gas density
fluctuations, between the Early and Late stages, which suggest that
the ISM becomes more homogeneous as the magnetic field grows.
As shown in Fig.~\ref{fig:nhot},
the average number density of the hot gas increases by a factor of two from the Early to the
Late stage. This enhances the cooling rate but cannot account fully for the reduction in the fractional volume of the hot gas by a factor of more than five.

\begin{figure}
\centering
	\includegraphics[width=0.88\columnwidth]{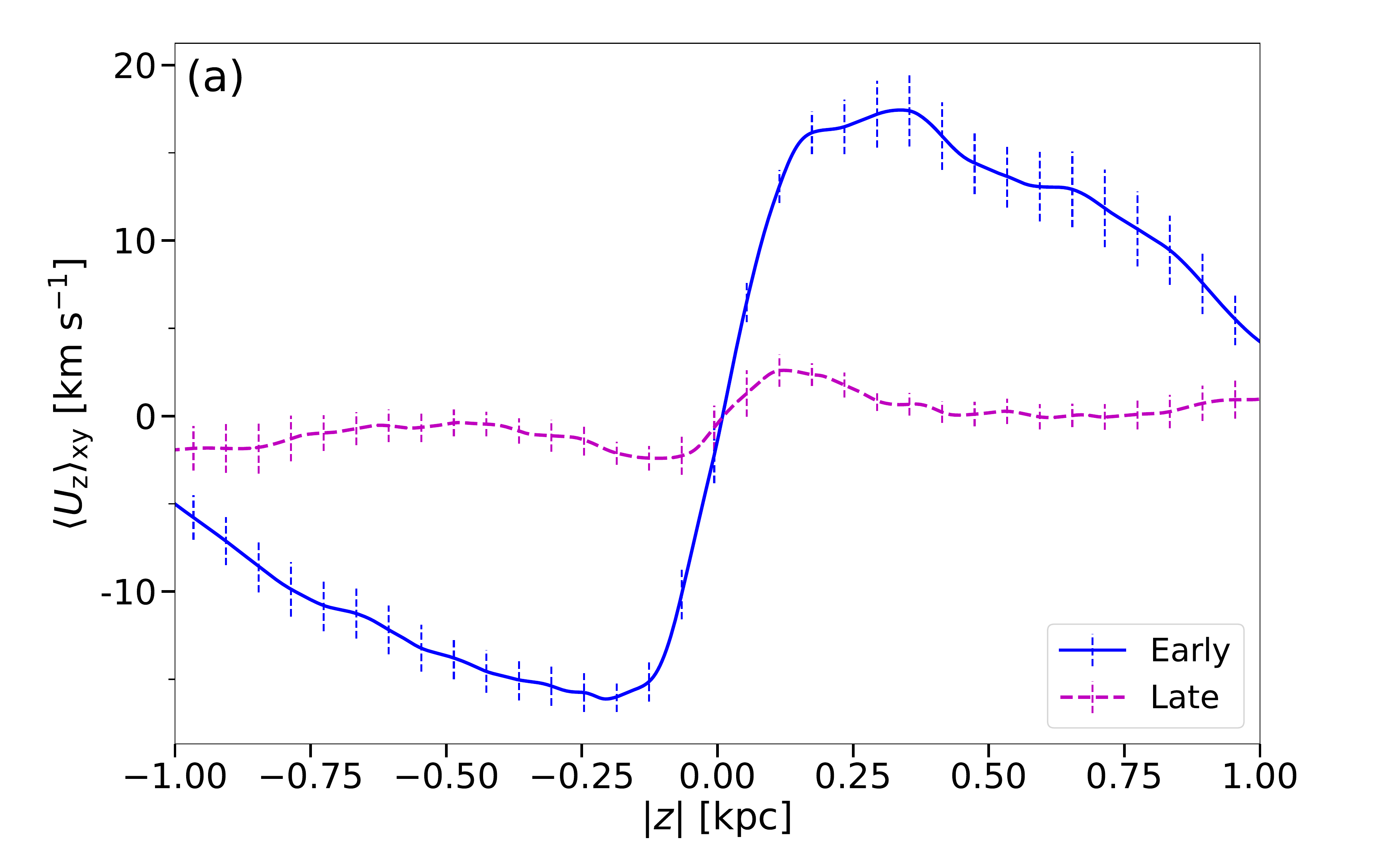}
	\includegraphics[width=0.85\columnwidth]{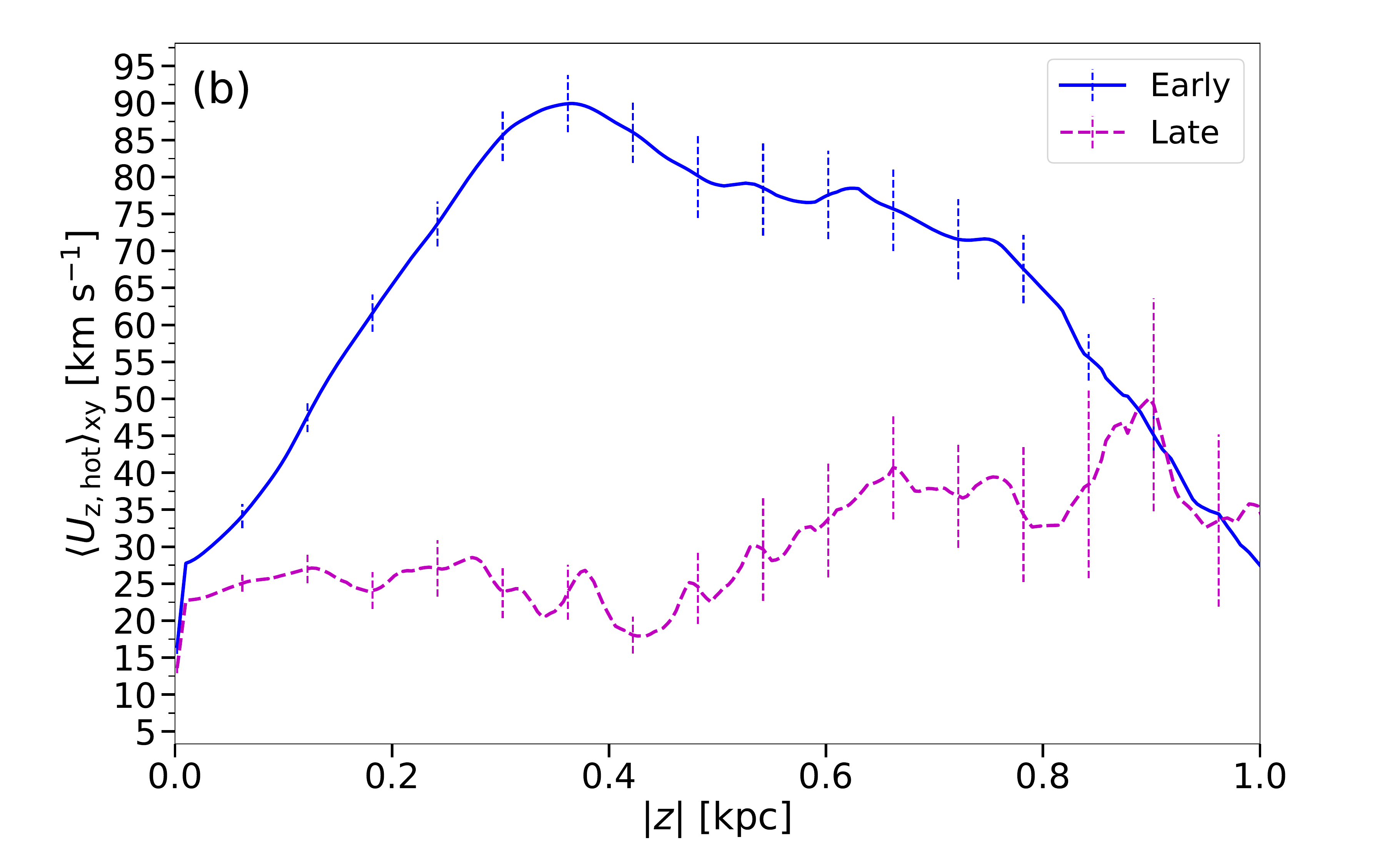}
	\caption{\textbf{(a)}~The {horizontally averaged} vertical velocity versus $|z|$ in the Early (solid, blue) and Late (dashed, magenta) stages of magnetic 
			field evolution \textbf{(b)}~As in panel (a) but for the hot gas alone.}
	\label{fig:vz_zavg}
\end{figure}

\begin{figure}
\centering
	\includegraphics[width=0.8\linewidth]{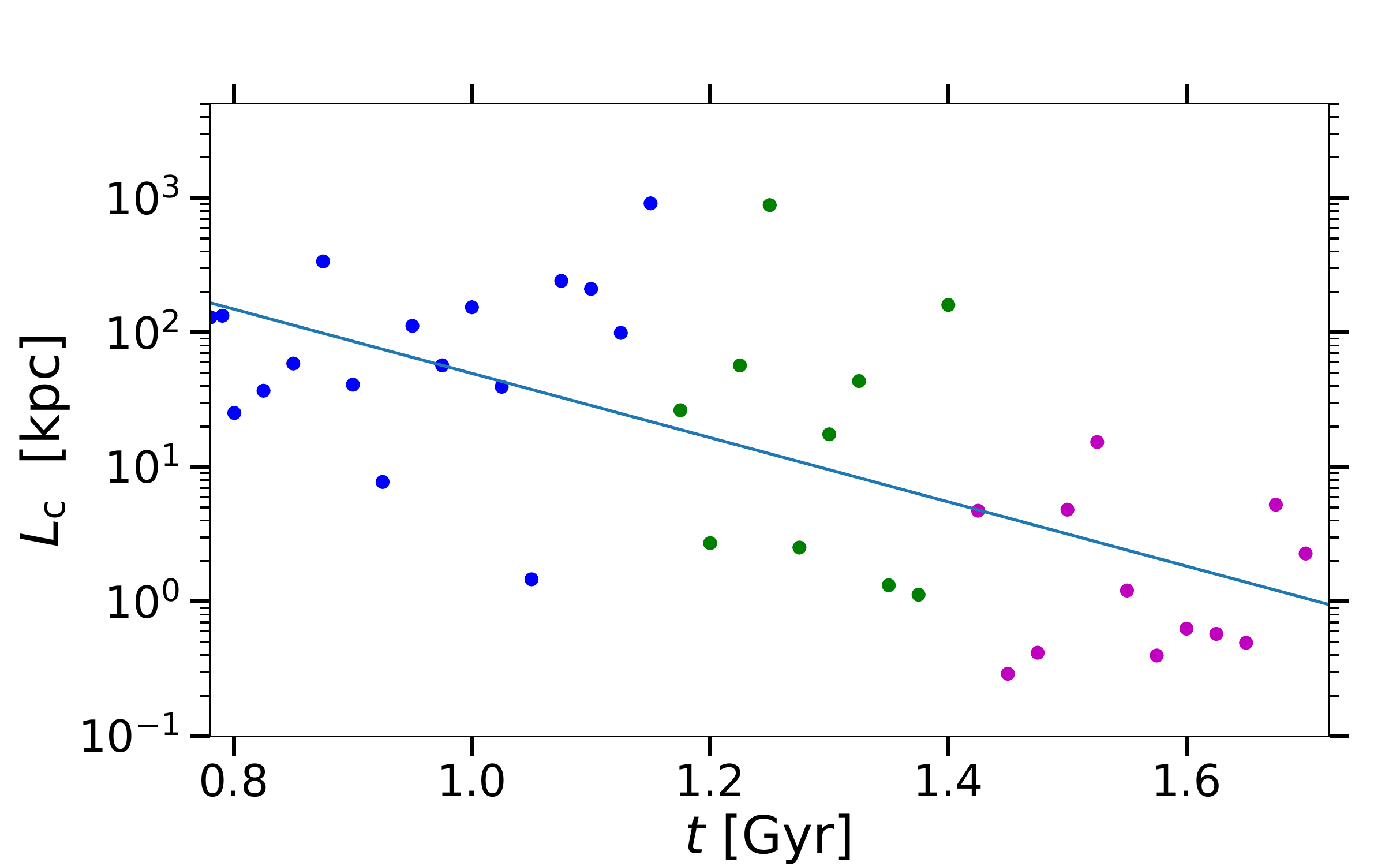}
	\caption{The evolution of the cooling length of the hot gas.
		The fitted line is obtained by linear regression of $\log L_{\mathrm{c}}$ against $t$. The fit is given by $L_{\mathrm{c}} = 10^{4.08-2.39(t/1~\mathrm{Gyr})}$.}
	\label{fig:cooling_length}
\end{figure}

The cooling of the hot gas is further enhanced by its longer residence time near the midplane, as its outflow is quenched by the magnetic field.
This is reflected in the decrease of cooling
length of hot gas, which is defined as $L_\text{c}=\tau_\text{c}U\hot,$ where $U\hot$ is the mean vertical velocity of 
the hot gas shown in Figure~\ref{fig:vz_zavg}\,\textbf{(b)} and  
$\tau_\text{c}=c_V T/(\rho\Lambda)$ is the radiative cooling time, with 
$\Lambda$ the cooling function 
\citep[described in detail in][]{GSFSM13}, $c_V$ is the specific heat and 
$T$ and $\rho$ are the gas temperature and density.
Figure~\ref{fig:cooling_length} shows that the cooling length decreases 
from $50$--$100$~kpc in the Early stage, to about $1$~kpc.
The magnetic fields affect the abundance of hot gas in two indirect ways:
firstly, it enhances the cooling rate of the hotter gas by increasing its density;
secondly, it opposes the outflow of the hot gas from the midplane, allowing it to cool
for a longer time.

\subsection{Magnetic quenching of vertical velocity}
\begin{figure*}
\centering
	\includegraphics[width=0.85\linewidth]{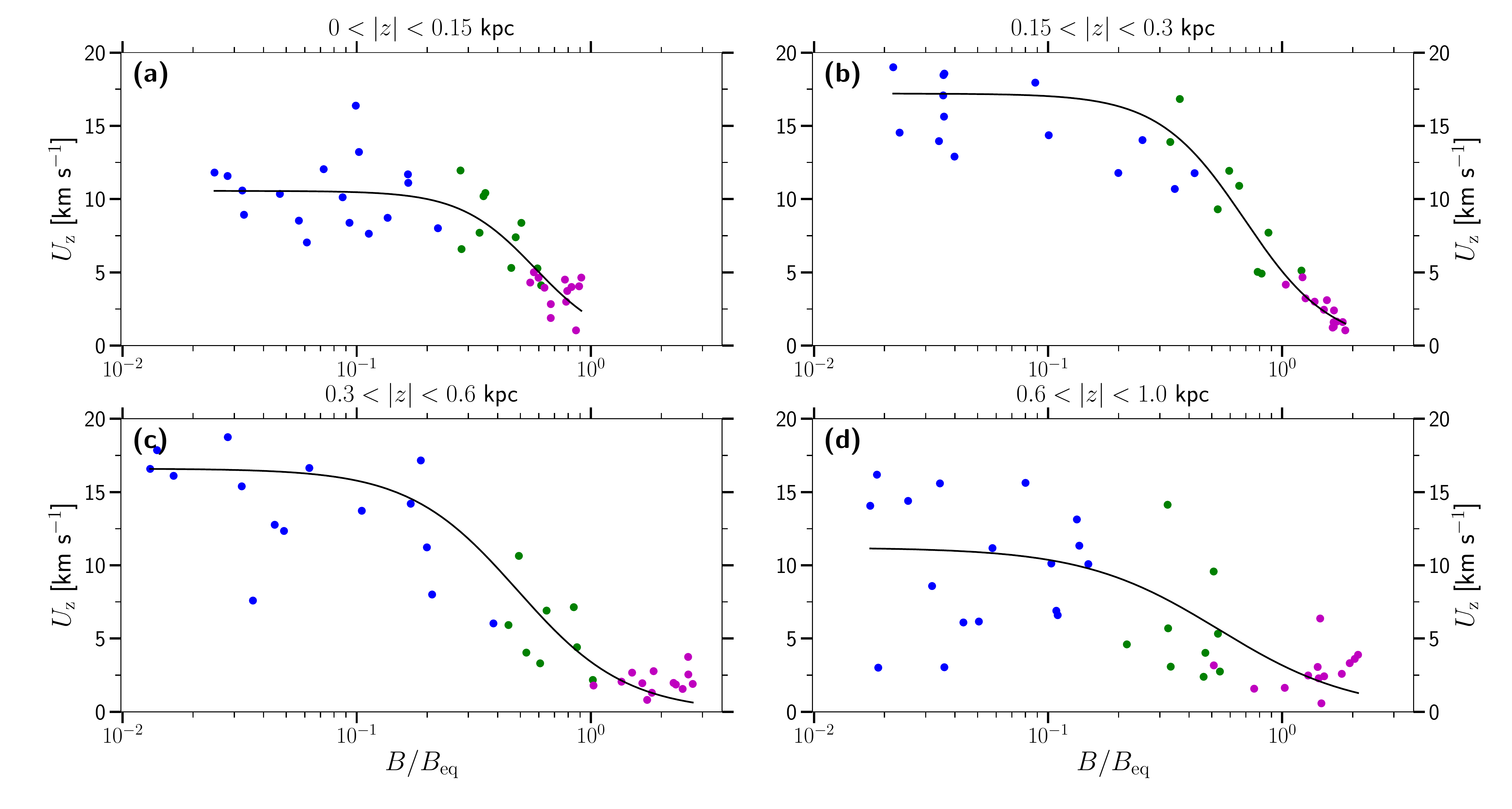}
\caption{The dependence of the mean vertical speed on the 
strength of the mean magnetic field {at} various distances from the mid-plane: 
	\textbf{(a)}~$0\leq|z|\leq0.15\kpc$, 
	\textbf{(b)}~$0.15\leq|z|\leq0.30\kpc$,
	\textbf{(c)}~$0.3\leq|z|\leq0.6\kpc$ and 
	\textbf{(d)}~$0.6\leq|z|\leq1.0\kpc$. The data points represent 
	horizontally averaged values of the mean vertical speed in 
	individual snapshots in the Early (blue), Transitional (green) and Late 
	(purple) stages of magnetic field evolution. Parameters of the fits 
	\eqref{uzB}, shown in black solid lines, are given in 
	Table~\ref{tab:uzB_fits}.}
	\label{fig:vz_zavg_b}
\end{figure*}

There are contradictory opinions about the effects of magnetic fields
in galactic outflows.
\citet{Boulares88}, \citet{BC90}, \citet{PV-SP95}, and \citet{BGE15} suggest that a large-scale 
magnetic field does affect vertical gas motions. However, the numerical simulations 
of \citet{Avillez05}, \citet{HJMBM12} and \citet{SILCC2} find no evidence for 
such an effect with an imposed plane-parallel magnetic field. 
To determine whether the decrease in vertical velocity is connected to the
magnetic field, we examine the relationship between
the dependence of the mean vertical velocity $|\langle U_z\rangle_{xy}|$ on mean
magnetic field strength\footnote{The total magnetic field is decomposed into mean and
random components using horizontal averaging. We also use horizontal averaging to calculate vertical profiles of quantities presented in this Paper.}, which is shown in Figure~\ref{fig:vz_zavg_b}.
It is useful to compare with Figure~6 of \citet{BGE15}.
Our results can be approximated by
\begin{equation}\label{uzB}
|\rsub{U}{z}|\approx 
\frac{U_0}{1+|\xi\langle B\rangle_{xy}/\langle B_{\mathrm{eq}}\rangle_{xy}|^n}\,,
\end{equation}
where $B_{\mathrm{eq}}$, {a function of $z$,} is the local equipartition magnetic field 
strength, $B_{\mathrm{eq}}^2=4\pi\rho u^2$, where $u$ is the turbulent gas velocity. The fitted  values for $U_0$, $\xi$, and $n$ are given in 
Table~\ref{tab:uzB_fits}. {\citet{BGE15} show a similar relation between 
$|\langle U_z\rangle|$ and $\langle B\rangle_{xy}/\langle B_{\mathrm{eq}}\rangle_{xy}$ (Figure~6). While fitted parameters differ,
both fits feature a steep decrease in vertical velocity for values of this ratio
above a threshold value. We find that this threshold value is $0.1-0.2$, in agreement with
\citet{BGE15}.
\begin{table}
 \begin{center}
 \caption{ Fits to the outflow speed, of the form \eqref{uzB}, at  various distances $|z|$ from 
 the mid-plane.}
 \label{tab:uzB_fits}
 {\scriptsize
    \begin{tabular}{lccc}
\hline
Distance to the mid-plane [kpc]         &$U_0$ [km\,s$^{-1}$]  &$\xi$ &$n$\\
\hline
\phantom{$0.15<$} $|z|<0.15$ 		& 11 & 1.7 & 2.7 \\
$0.15<|z|<0.3$	    & 17 & 1.4 & 2.4 \\
\phantom{0}$0.3<|z|<0.6$	    & 17 & 2.1 & 1.9 \\
\phantom{0}$0.6<|z|<1.0$	    & 11 & 1.9 & 1.5 \\
\phantom{2cm}$|z|=0.8$~\citep{BGE15}    	    & 12 & 1.2 & 2 \\
\hline
    \end{tabular}
  }
 \end{center}
\end{table}

\section{Gas pressure}\label{sec:gas_pressure}

Figure~\ref{fig:zavg_pressure} shows vertical profiles of various pressure components
for the Early and Late stages.
In the Early stage, thermal and kinetic pressure are dominant; while magnetic pressure
is approximately an order of magnitude smaller.
The total pressure gradient is constrained by the weight of the gas and the strength of 
the gravity field. The latter does not change during the simulation, so the 
increase over time in the vertical gradient of the total pressure 
is due to the ejection of gas to larger altitudes. The contributions of the thermal and turbulent pressures decrease 
together with the total pressure as magnetic field grows (the relative 
contribution of the turbulent pressure decreases especially strongly in the 
Late stage) as they are replaced by the magnetic pressure. The contributions of the
gradients of the various pressure components to the force balance are discussed in Section~\ref{sec:force}.

Remarkably, magnetic pressure is the only part 
of the total pressure that varies with $|z|$ non-monotonically, confining the 
gas at $|z|\lesssim0.3\kpc$ and producing an outwardly directed force above 
that level.
%
\begin{figure}
\centering
\includegraphics[width=0.99\linewidth]{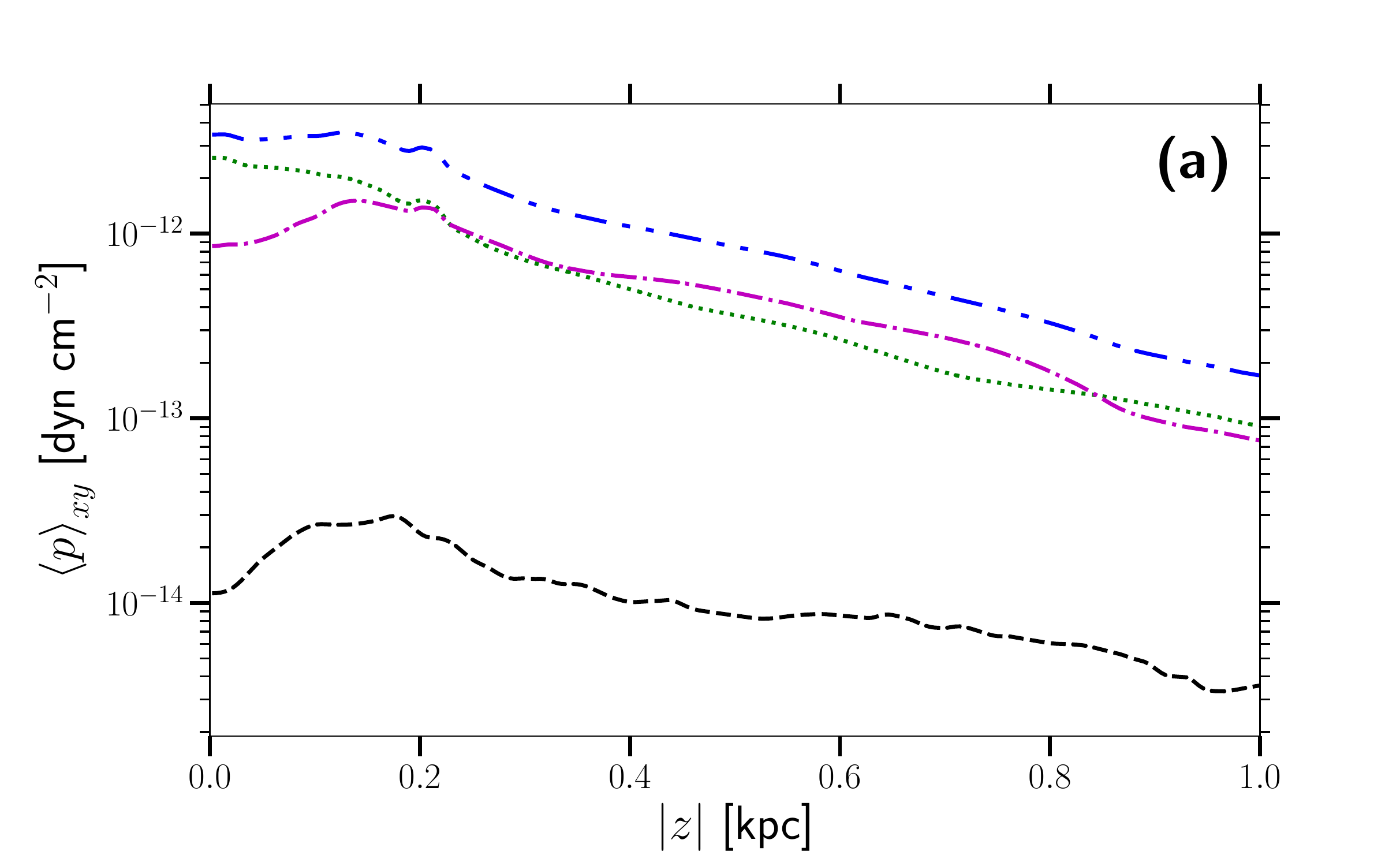}
\includegraphics[width=0.99\linewidth]{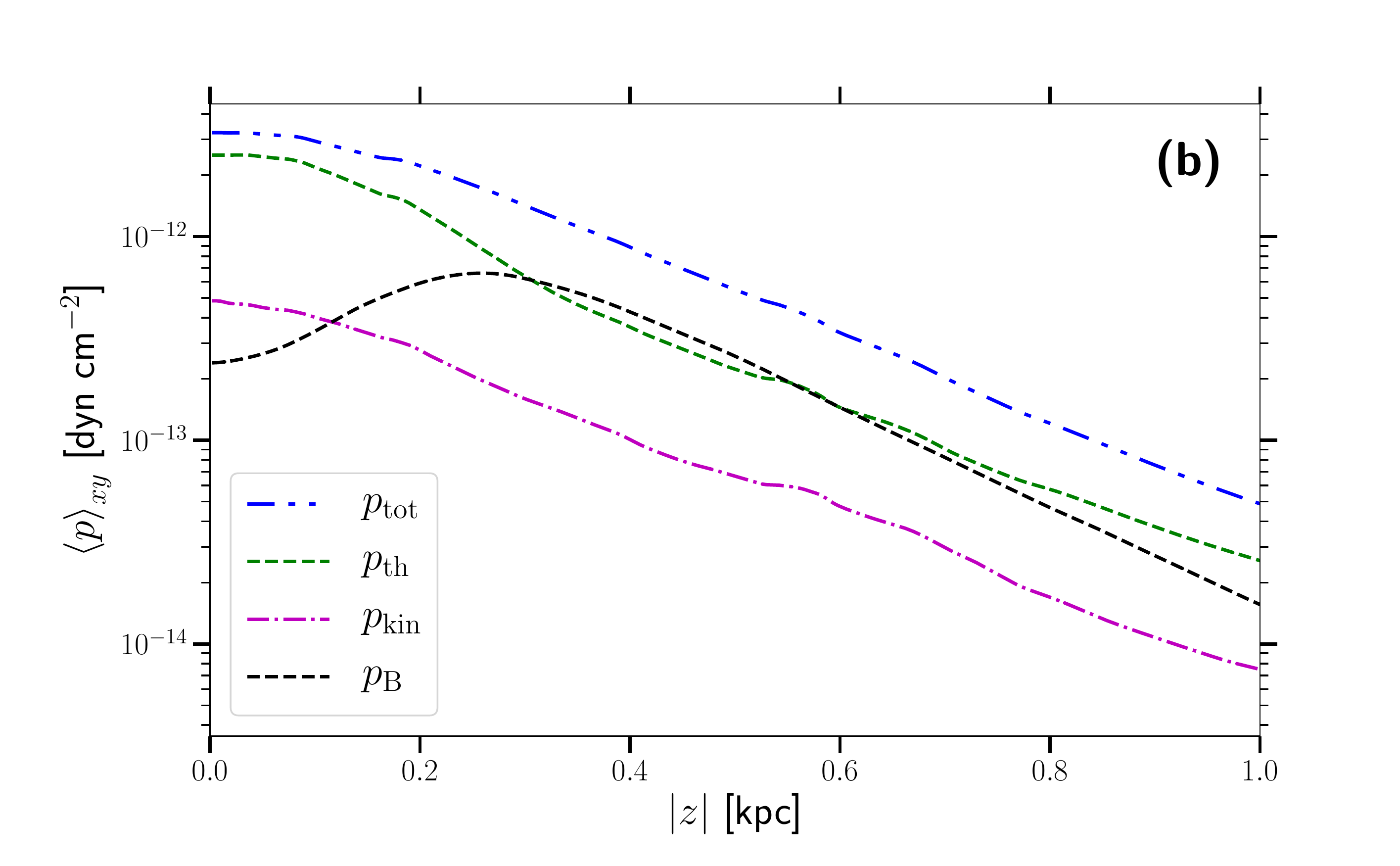}
\caption{Vertical profiles of the total pressure, $p_\text{tot}=p_\text{th}+p_\text{kin}+p_\text{B}$, thermal pressure 
$p_\text{th}$, turbulent pressure $p_\text{kin}$, and 
magnetic pressure $p_\text{B}$, and their variation as the system evolves from the {\textbf{(a)}}~Early
to {\textbf{(b)}}~Late stage.
}
\label{fig:zavg_pressure}
\end{figure}
In the Late stage, magnetic pressure is within an order of magnitude of thermal and
kinetic pressure within 200~pc of the midplane. It is in equipartition with 
thermal pressure, and in supra-equipartition with kinetic pressure further
away from the midplane. Stronger magnetic field may be produced near the midplane
and advected to larger altitudes to dominate over the random flows there, as we discuss in Section~\ref{sec:bz}.

\section{Vertical force balance}\label{sec:force}
To understand the dynamics of the ISM, in particular the effects of magnetic 
tension and pressure gradient on the vertical flow, we consider the averaged 
vertical momentum equation
\begin{align}\label{eqn:vrt_bal}
\deriv{}{t}\xyavg{\rho U_z}	&=\xyavg{\rho  g_z} - \deriv{P}{z}
					+\frac{1}{4\pi}\xyavg{(\vect{B}\cdot\nabla)B_z}
					+\mathcal{D},
\intertext{where the total mean pressure is}
P&=\xyavg{p_{\mathrm{th}}} + \frac{1}{8\pi}\xyavg{|\bvec{B}|^2}
+\xyavg{\rho U^2_{z}},
\end{align}
$p_\mathrm{th}$ is thermal gas pressure, $\rho$ is the gas mass density,
$g_z$ is the vertical gravitational acceleration, and $\mathcal{D}$ represents
diffusion terms, including numerical diffusion.
We compute and compare individual terms in this equation but, since the 
physical diffusion is just one contribution to the momentum dissipation, we do 
not include the diffusion term in the analysis that follows. This
should be kept in mind when discussing the force balance but this does not 
affect our conclusions regarding the relative magnitudes of various terms.
{The derivation of this equation, in absence of diffusion, is provided in Appendix~\ref{deriv}.}

\begin{figure*}
\includegraphics[width=0.95\linewidth]{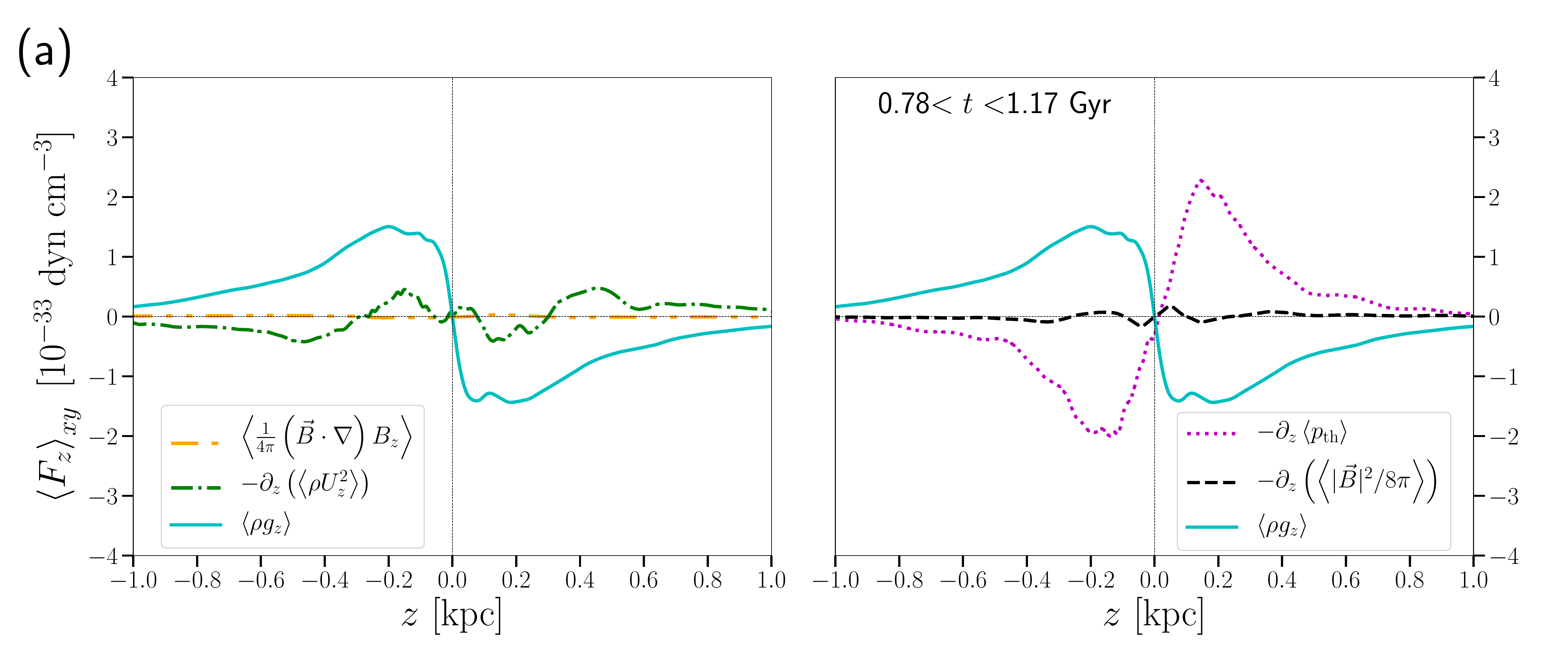}
\includegraphics[width=0.95\linewidth]{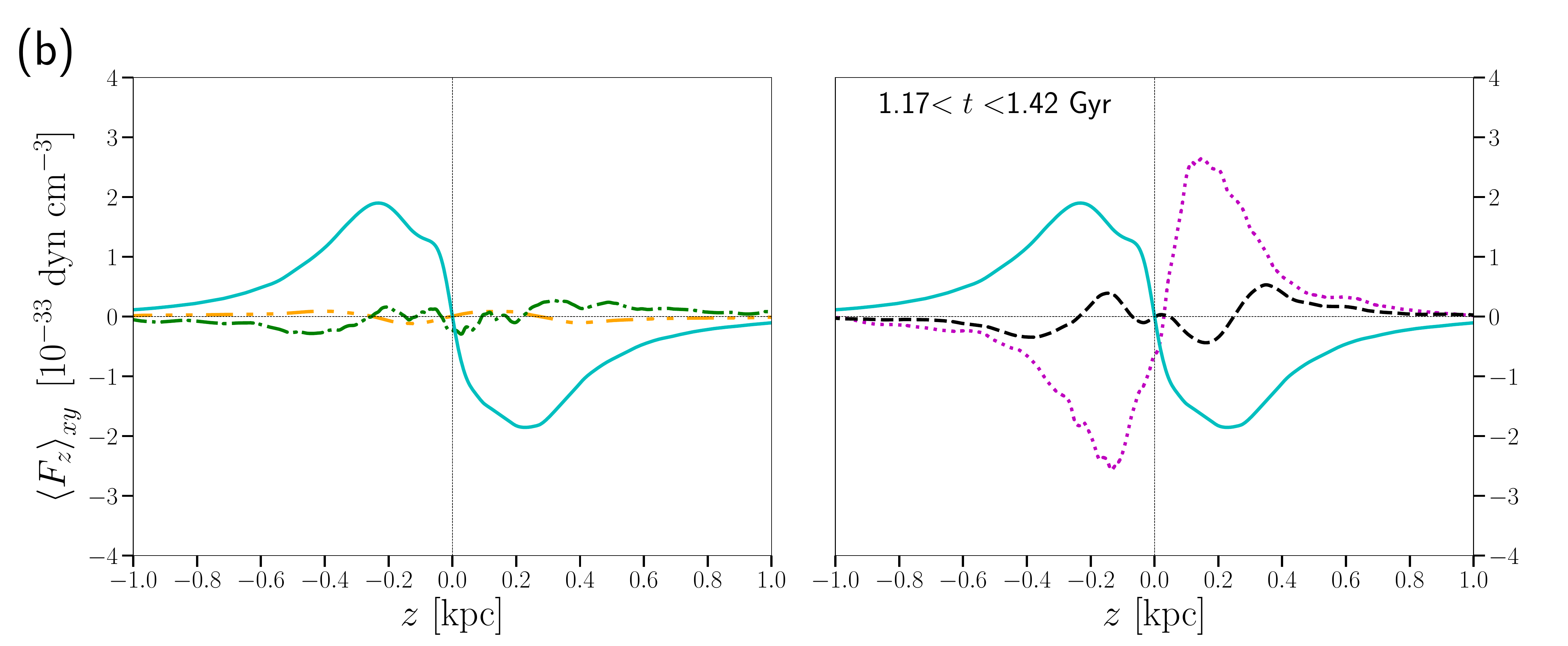}
\includegraphics[width=0.95\linewidth]{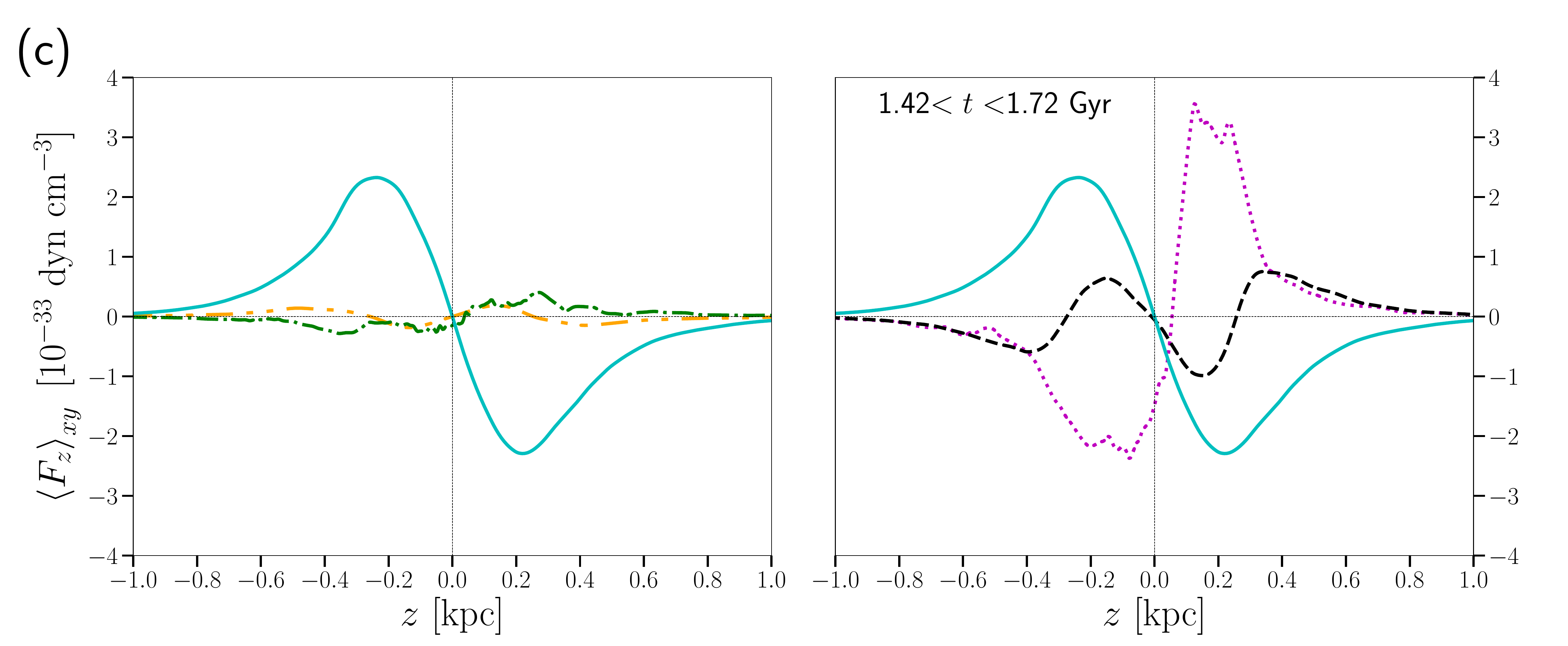}
\caption{Horizontal averages of the individual terms in the vertical momentum 
equation~\eqref{eqn:vrt_bal} in the (\textbf{a})~Early, 
(\textbf{b})~Transitional and (\textbf{c})~Late stages, from top to bottom. The 
vertical profile of the gravity force (solid, light-blue) is shown in all panels 
for reference, thermal and magnetic pressure gradients are shown dotted 
(magenta) and dashed (black), respectively, and magnetic tension and turbulent 
pressure gradient are shown solid (blue) and dash-dotted (green), respectively.}
\label{fig:mom_trms}
\end{figure*}

Figure~\ref{fig:mom_trms} show the vertical profiles of the individual terms in 
the vertical momentum equation \eqref{eqn:vrt_bal} for the Early (upper row), 
Transitional (middle) and Late (bottom) stages, respectively. The contribution
of gravity is shown in all panels with solid blue for reference. The 
magnetic 
pressure gradient and tension are negligible in the Early stage, 
Fig.~\ref{fig:mom_trms}\,\textbf{a}  when the magnetic field is still
growing exponentially and has minor dynamical significance.
The thermal pressure gradient (dotted, magenta) is close to 
balance with the gravity force, exceeding it most notably   
around $|z|=0.2\kpc$ to drive the systematic gas outflow. The turbulent pressure gradient 
(dash-dotted, green) is subdominant. This picture remains qualitatively similar 
in the Transitional stage but the magnetic pressure gradient (dashed, black) 
already makes a noticeable contribution to the force balance.
The importance of 
magnetic pressure increases in the Late stage where it assists gravity to 
confine the gas layer at $|z|<0.3\kpc$, but combines with
thermal pressure against gravity at larger distances from the mid-plane. Horizontally averaged magnetic tension 
(solid, blue) opposes the magnetic pressure gradient but this force is 
subdominant at all times and all altitudes. The turbulent pressure gradient is 
also weak.

Altogether, the vertical force balance is dominated by gravity and the thermal 
pressure gradient. Contrary to the suggestion of \citet{BC90}, magnetic tension 
is negligible within $|z|\lesssim1\kpc$ and we expect it is unlikely to become 
more important at larger altitudes.

\section{Vertical distribution of the mean magnetic field}\label{sec:bz}
A notable and unexpected feature of the distribution of the magnetic field is that it has its maximum away from the midplane
as shown in Fig.~\ref{fig:Bbuz}. In the Late stage,
the maximum of the mean field strength is
located around $|z|\simeq250$\,pc, while for the random field it is slightly
nearer the midplane at $|z|\simeq200$\,pc.

\begin{figure}
	\centering
	\includegraphics[width=0.8\linewidth]{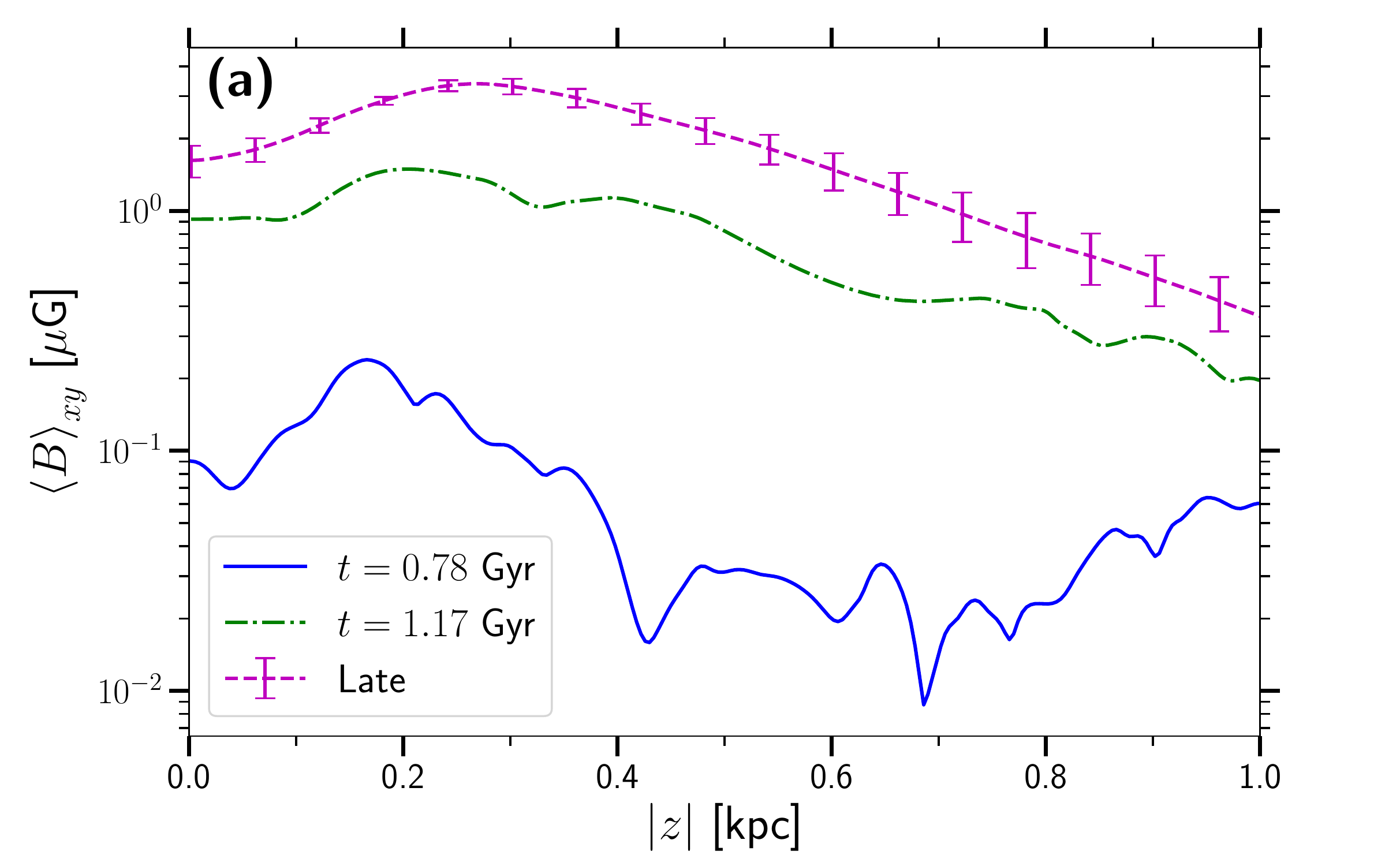}
	\includegraphics[width=0.8\linewidth]{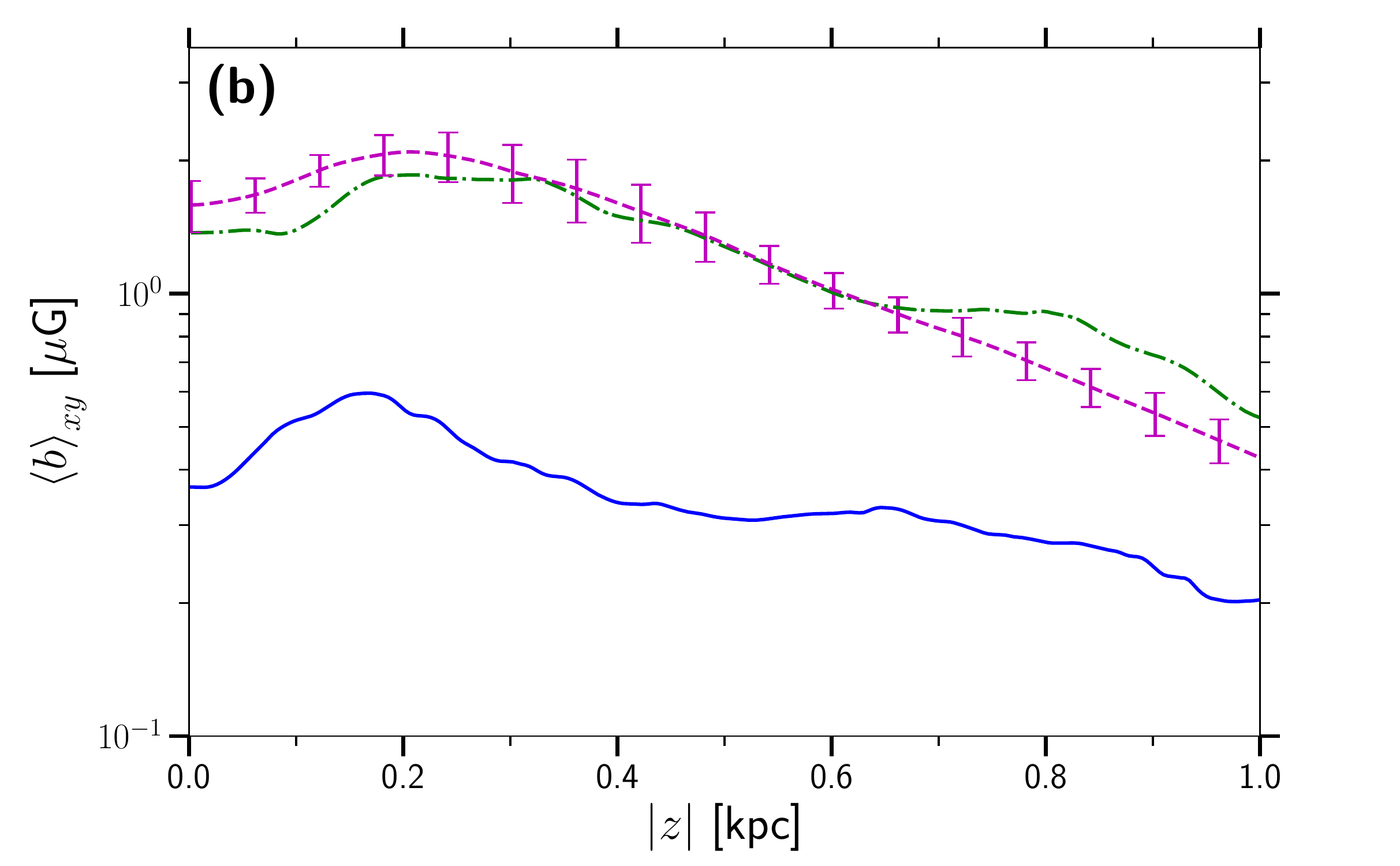}
	\caption{\label{fig:Bbuz}The vertical profiles of \textbf{(a)}~mean and 
	\textbf{(b)}~random magnetic field strength in the Early (blue, 
	dashed) and Late (magenta, solid) stages.}
\end{figure}

The mean magnetic field can be redistributed from the 
mid-plane by turbulent diamagnetism, i.e., the 
transport of the mean magnetic field from a region
with enhanced turbulence at an effective speed of 
$-\tfrac12\nabla\eta\turb$, where 
\begin{align*}
\eta\turb\simeq\frac{1}{3} \tau u_0^2
           \simeq\frac{1}{3} u_0l_0,
\end{align*}
is the turbulent magnetic diffusivity,
$\tau$ and $l_0$ the correlation time and length of the flow, $u_0$ its rms speed \citep{Z56,RS75}.
The correlation length, $l_0$, of the random velocity in the warm 
gas increases from about 70 to 100 pc between $|z|=0$ and $0.4\kpc$, 
whereas $u_0$ decreases from $13$ to $4\kms$, over the distance 
$\Delta z=0.4\kpc$ \citep[Table~3 of][]{HSSFG17}.
Thus, 
\[-\frac{1}{2}\nabla\eta\turb\simeq-\frac{1}{6}\frac{\Delta(l_0u_0)}{\Delta z}\simeq0.2\kms,\]
where $\Delta$ denotes the increment in the corresponding variable.
 {With these estimates, the maximum of the mean magnetic field would
be displaced {away from the mid-plane by a distance of}
$-\tfrac12 t\nabla\eta\turb\simeq120\pc$ over a period of} $0.6\Gyr$.
While this {cannot fully account for the location of maximal field strength}
in Fig~\ref{fig:Bbuz}, turbulent diamagnetism {probably} contributes to the 
transport of mean magnetic field. The magnetic pressure exceeds the turbulent (kinetic) pressure at $|z| > 0.2\kpc$ (as shown in Fig.~\ref{fig:zavg_pressure}), which also
suggests that magnetic field is systematically transported away from its generation region.

Since a significant part of the random field is generated by tangling of the mean field by the random flow, 
it is understandable that both have a maximum away from $z=0$~kpc.

Another factor that could contribute to the non-monotonic variation of the
magnetic field strength with $|z|$ is a similarly non-monotonic distribution of 
intensity of dynamo action, quantified by the dynamo number.
However, careful assessment of this possibility requires sophisticated dedicated analysis,
which is beyond the scope of this paper.

As mentioned above, we do not suggest that 
the vertical profiles shown in Figures~\ref{fig:B2O_rho} and \ref{fig:Bbuz} 
necessarily occur in the Milky Way or any other specific galaxy but rather 
argue that this behaviour is physically meaningful and that its possibility 
should be kept in mind in the interpretation of observations and the assessment 
of various effects of interstellar magnetic fields.

\section{Conclusions}\label{sec:Conclusions}
We have found a significant qualitative difference between magnetic
effects in our model (in agreement with the results of \cite{BGE15}) and those models adopting strong
\textit{imposed} magnetic fields. Any numerical model of this kind must provide an
initial value and spatial configuration for the magnetic field. If a strong imposed
field is used, then it is crucial to use a spatial configuration which reflects the
configuration of the evolved magnetic field.

We find that systematic outflow speed is reduced throughout the disc and disc-halo
interface as the magnetic field grows, approximated by Eq.~\eqref{uzB}. Below $|z|=300\pc$ the magnetic pressure gradient opposes the outflow of gas
driven by thermal pressure. 
We, and \citet{BGE15}, find that
there is a rapid decrease in $\rsub{U}{z}$ for $\rsub{B}{h}/\rsub{B}{eq}>0.1-0.3$. Models with imposed magnetic fields do not
capture this effect, which is vital to modelling how outflows feed the galactic halo.

Magnetization of the simulated ISM leads to an increase in the density of hot gas,
particularly close to the midplane where hot gas is produced
by SNe. However, we also find that fractional volume of hot gas decreases
by up to an order of magnitude from the Early to the Late stage, due to an increased cooling rate and a reduced outflow speed of the hot gas. This is
evident in the visualisation shown in Fig.~\ref{fig:sn_fx}, with the hot gas
structures becoming more compact as the magnetic field grows.

The changes outlined above combine to produce a qualitative change in the
vertical density profile of ISM gas between the Early and Late stages. Once the magnetic field has reached a steady state, the density profile is flatter within 300~pc of the midplane. Above this height the density decreases more rapidly with distance from the midplane compared to the Early stage.

\section*{Acknowledgements}
FAG acknowledges financial support of HPC-EUROPA2, Project No.~228398,
and the Academy of Finland's Project 272157. This work has benefited from 
access to the resources of the Grand Challenge Project SNDYN of the CSC-IT 
Center for Science Ltd., Finland. AS, AF and PB were supported by the 
Leverhulme Trust Grant RPG-2014-427 and STFC Grant ST/N000900/1 (Project 2). 

\bibliographystyle{mn2e}
  \bibliography{refs}
\appendix
\section{The momentum equation}\label{deriv}
 The $z$-component of the Navier-Stokes equation, neglecting dissipation, and the continuity equation are given by
  \begin{align}
    \partial_t U_z&=-\left(\bvec{U}\cdot\nabla\right)U_z-\frac{1}{\rho}\partial_zp_{\mathrm{th}}
    +\frac{1}{4\pi\rho}\left[\bvec{j}\times\bvec{B}\right]_z+ g_z\label{eqn:nsz}\\
    \partial_t\rho&=-\nabla\cdot\left(\rho\bvec{U}\right)\label{eqn:cty},
  \end{align}
  where vertical gravitational acceleration, $g_z$, due to
  stellar and dark halo matter follows 
  \cite{Kuijken89}, and
  \begin{equation}
    \frac{1}{4\pi}\bvec{j}\times\bvec{B}=\frac{1}{4\pi}(\bvec{B}\cdot\nabla)\bvec{B}-\nabla\left(\frac{|\bvec{B}|^2}{8\pi}\right),
    \label{eqn:mag_dcmp}
  \end{equation} 
  Equations~\ref{eqn:nsz} and \ref{eqn:cty} can be combined into
  \begin{align*}
    \partial_t\left(\rho U_z\right)&=-\left[\rho\left(\bvec{U}\cdot\nabla\right)U_z+U_z\nabla\cdot\left(\rho\bvec{U}\right)\right]
   -\partial_zp_{\mathrm{th}}\\&+\frac{1}{4\pi}\left[\bvec{j}\times\bvec{B}\right]_z+\rho g_z.
  \end{align*}
  \noindent We then use the identity
  \[
    \nabla\cdot\left(\rho U_z\bvec{U}\right)=\rho\left(\bvec{U}\cdot\nabla\right)U_z+U_z\nabla\cdot\left(\rho\bvec{U}\right), 
  \]
  \noindent to obtain
  \[
    \partial_t\left(\rho U_z\right)=-\nabla\cdot\left(\rho U_z\bvec{U}\right)
   -\partial_zp_{\mathrm{th}}+\frac{1}{4\pi}\left[\bvec{j}\times\bvec{B}\right]_z+\rho g_z.
  \]
  \noindent Now, we substitute the $z$-component of Equation~\ref{eqn:mag_dcmp} into this equation:
  \begin{align}
    \partial_t\left(\rho U_z\right)=&
    \rho g_z-\partial_z\left(p_{\mathrm{th}}+\frac{|\bvec{B}|^2}{8\pi}\right)\nonumber\\
    &+\frac{1}{4\pi}(\bvec{B}\cdot\nabla)B_z-\nabla\cdot\left(\rho U_z\bvec{U}\right)\label{eqn:int_eq1}
  \end{align} 
\noindent This equation is horizontally averaged to give
  \begin{align*}
    \partial_t\left(\left<\rho U_z\right>\right)=&
    \left<\rho g_z\right>-\partial_z\left(\left<p_{\mathrm{th}}\right>+\left<\frac{|\bvec{B}|^2}{8\pi}\right>\right)\\
    &+\left<\frac{1}{4\pi}(\bvec{B}\cdot\nabla)B_z\right>-\left<\nabla\cdot\left(\rho U_z\bvec{U}\right)\right>.
  \end{align*}
  \noindent The following term is expanded, which gives
  \begin{align*}
    \left<\nabla\cdot\left(\rho U_z\bvec{U}\right)\right>=&\left<\partial_x\left(\rho U_zU_x\right)\right>
    +\left<\partial_y\left(\rho U_zU_y\right)\right>\\
    &+\left<\partial_z\left(\rho U^2_z\right)\right>.
  \end{align*}
  \noindent However, horizontal averages of $x$ and $y$ derivatives are zero
  due to periodic boundary conditions in $x$ and $y$. Thus, we have
  \[
    \left<\nabla\cdot\left(\rho U_z\bvec{U}\right)\right>=
    \partial_z\left(\left<\rho U^2_z\right>\right).
  \]
  Substituting this term into Equation~\ref{eqn:int_eq1} produces
  \begin{align}
    \partial_t\left(\left<\rho U_z\right>\right)=&
    \left<\rho g_z\right>-\partial_z\left(\left<p_{\mathrm{th}}\right>+\left<\frac{|\bvec{B}|^2}{8\pi}\right>\right)\nonumber\\
    &+\left<\frac{1}{4\pi}(\bvec{B}\cdot\nabla)B_z\right>-\partial_z\left(\left<\rho U^2_z\right>\right).
    \label{eqn:vrt_bal1}
  \end{align}
  \noindent 

\label{lastpage}

\end{document}